# Model-based analysis of brain activity reveals the hierarchy of language in 305 subjects

Charlotte Caucheteux[1,2], Alexandre Gramfort[2], Jean-Rémi King[1,3]

[1]Facebook AI Research, Paris, France
[2]Université Paris-Saclay, Inria, CEA, Palaiseau, France
[3]École normale supérieure, PSL University, CNRS, Paris, France

## Abstract

A popular approach to decompose the neural bases of language consists in correlating, across individuals, the brain responses to different stimuli (e.g. regular speech versus scrambled words, sentences, or paragraphs). Although successful, this 'model-free' approach necessitates the acquisition of a large and costly set of neuroimaging data. Here, we show that a model-based approach can reach equivalent results within subjects exposed to natural stimuli. We capitalize on the recently-discovered similarities between deep language models and the human brain to compute the mapping between i) the brain responses to *regular* speech and ii) the activations of deep language models elicited by *modified* stimuli (e.g. scrambled words, sentences, or paragraphs). Our model-based approach successfully replicates the seminal study of (Lerner et al., 2011), which revealed the hierarchy of language areas by comparing the functional-magnetic resonance imaging (fMRI) of seven subjects listening to 7 min of both regular and scrambled narratives. We further extend and precise these results to the brain signals of 305 individuals listening to 4.1 hours of narrated stories. Overall, this study paves the way for efficient and flexible analyses of the brain bases of language.

## 1 Introduction

One of the most successful paradigms to decompose the brain bases of language consists in correlating the brain responses of multiple subjects listening to the same carefully controlled stimuli (Brennan et al., 2012; Fedorenko et al., 2016; Blank et al., 2016; Mollica et al., 2019). In particular, (Lerner et al., 2011) recorded subjects with functional magnetic resonance imaging (fMRI) while they listened to a story whose (1) sounds (2) words, (3) sentences or (4) paragraphs were scrambled, as well as (5) to the regular version of the story (Figure 1A). The authors then estimated the Inter Subject Correlation (ISC), *i.e.,* the correlation between i) the brain activity of a voxel in response to one scrambling condition and ii) the brain activity of a voxel averaged across all other subjects, in response to the same scrambled stimulus (Figure 1B). While successful, this 'model-free' approach is costly: it requires $n_\text{subjects} \times n_\text{conditions}$ acquisitions of brain activity in response to the same variably scrambled stimuli.

Here, we investigate whether and how a model-based approach can replicate Lerner et al.'s findings, even if we only have access to the recordings elicited by the regular story in a single subject. We further apply the method to extend Lerner et al's results to a large dataset of 305 individuals.

## 2 Methods

First, we formalize the 'model-free' and 'model-based' approaches in the context of narrative listening, and explicit the link between the two.

**Definitions** Let's define

- $w = $ ('Once', 'upon', ... , 'The', 'end.') the regular story. $\Omega$ the story's vocabulary.

- $w_{|\text{sound}}, w_{|\text{word}}, w_{|\text{sent}}, w_{|\text{parag}}$ the story scrambled at the acoustic, word, sentence and paragraph level, respectively, following the setting of Lerner et al. (cf. Appendix B for the scrambling paradigm).

- 🧠 $: \Omega^M \to \mathbb{R}^T$: the function returning the brain recordings of length $T$ time samples (*i.e.,* the number of fMRI pulses) induced by a sequence of $M$ words.

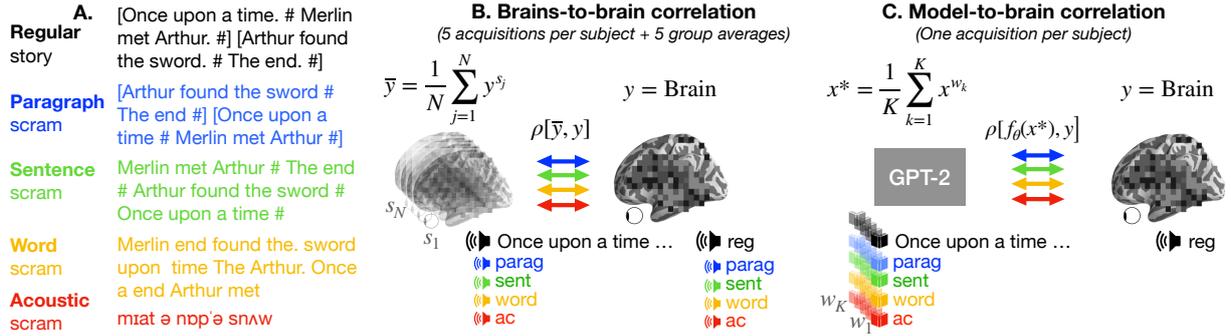

Figure 1: **Objective and methods A.** In Lerner et al.'s seminal study, each subject is presented successively with i) a 7 min long story (black), ii) the same story after its paragraphs (blue) iii) sentences iv) words (orange) or iv) acoustics (red) has been scrambled. **B.** For each condition, subject and voxel, the authors compute the inter-subject correlation (ISC), i.e the correlation $\rho$ between i) the brain of the current subject $y$ and ii) the average brain signals of the *other* subjects $\bar{y}$. This method allows to decompose the hierarchy of language processing in the brain, from the acoustic to the paragraph level. **C.** We aim to replicate the results of Lerner et al. using only the recordings induced by the regular story (black). To this aim, we scramble, not the stimulus of the subject, but the inputs of a deep language model (GPT-2). For each condition (word, sentence or paragraph), we extract the corresponding activations $x^*$ averaged over $K$ random scrambles. We then compare the brain signals of the current subject $y$ with the activations $x^*$ elicited by the scrambled texts, after a linear transformation $f_\theta$ that maps $x^*$ onto a brain-like space. Because GPT-2 is not trained to process waveform, we use the phonemes, stresses and tones of the stimulus instead of $x^*$ for the acoustic condition.

- 🤖 : $\Omega^M \to \mathbb{R}^{M \times D}$ the function returning the activations of a deep language model induced by a sequence of $M$ words.

- $y \in \mathbb{R}^T$ the brain recordings of one subject elicited by $w$, recorded at one voxel. Here, 🧠$(w) = y$.

- $y_{|\text{sound}}, y_{|\text{word}}, y_{|\text{sent}}, y_{|\text{parag}}$ the recordings elicited by the scrambled versions of $w$.

- $\rho : \mathbb{R}^T \times \mathbb{R}^T \to \mathbb{R}$, Pearson's correlation

For clarity, we describe below the model-free and model-based approaches for the *sentence* condition. The same methods can be used for the sound, word and paragraph conditions.

**Model-free analysis** Lerner et al. do not have a model of how the brain should react to sentences. Instead, they assume that the neural signature of sentence-level processing corresponds to the brain response shared across all subjects listening to scrambled sentences $w_{|\text{sent}}$. They thus compute the 'ISC score' for each subject, *i.e.*, the correlation between i) the brain response to the scrambled story $w_{|\text{sent}}$ of a given subject ($y_{|\text{sent}}$) and ii) the brain response to the same stimulus averaged across all other subjects ($\overline{y_{|\text{sent}}}$):

$$R = \rho\big(y_{|\text{sent}}, \overline{y_{|\text{sent}}}\big) \quad . \tag{1}$$

This approach boils down to a leave-one-subject-out cross-validation, using Pearson correlation as evaluation metric and the average population response as estimator.

**Model-based analysis** Here, we propose a model-based analysis to circumvent the need for a) the population average $\bar{y}$, b) the scrambled stimuli $w_{|\text{sent}}$.

To eliminate the need for the population average, we capitalize on the recent findings that deep language models tend to linearly predict brain responses to language (Jain and Huth, 2018; Gauthier and Levy, 2019; Toneva and Wehbe, 2019; Schrimpf et al., 2020; Caucheteux and King, 2020). We can thus assume that the average brain response ($\overline{🧠}$) can be well approximated by $f_\theta$, a linear function that maps the deep language model to the brain response. *i.e.,*

$$i) \quad \overline{🧠} \approx f_\theta \circ 🤖 \quad .$$

In practice, the coefficients $\theta$ of $f_\theta$ are estimated using ridge regression. Finite Impulse Response functions are employed to allow the activations of the deep language model of length $M$ (number of words) to map onto the slow and delayed brain recordings of length $T$ (number of pulses) (cf. Appendix C).

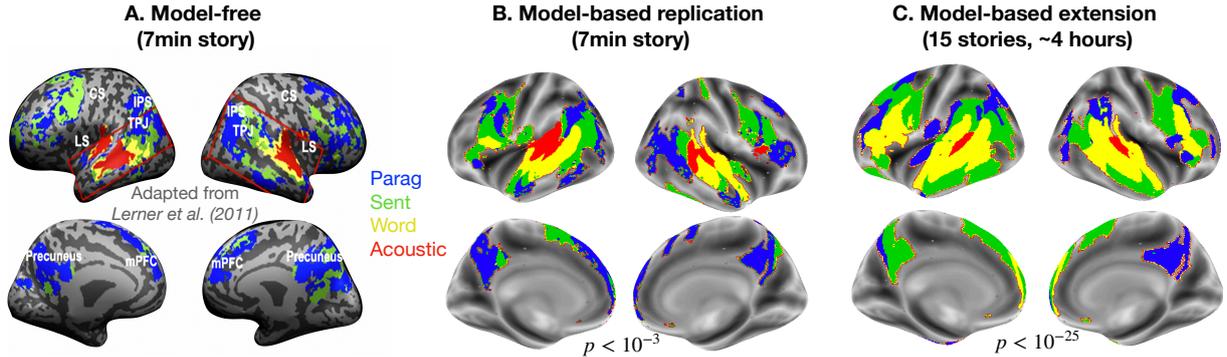

Figure 2: **Results.** Following Lerner et al's, a brain region is considered to process 'acoustic' level information if its acoustic score (either brains-to-brain or model-to-brain correlation) is significant (red). It is considered to process 'word'-level (yellow) if its word score is significant but not its acoustic one – and similarly for 'sentence' (green) and 'paragraph' (blue). **A.** Adapted from (Lerner et al., 2011). Labels are based on the brains-to-brain correlation scores (Figure 1B) averaged over seven subjects listening to a 7 min story. **B.** Labels are based on the model-to-brain scores (Figure 1C), averaged over 75 subjects listening to the same 7 min story. Significance is inferred using a Wilcoxon test across subjects, corrected with False Discovery Rate (FDR) across the 465 brain regions in each hemisphere (*cf.* Appendix D), with a significance threshold of $p < 10^{-3}$ (*cf.* Appendix E). **C.** Same as B., but on the brain of 305 subjects listening to 4 hours of 15 audio stories (including the 7 min one). Because of the large number of subjects, the significance threshold is set to $p < 10^{-25}$.

To eliminate the need for the scrambled stimuli, we show below that equation (1) can be rewritten only as a function of $w$ as opposed to $w_{|\text{sent}}$.

First, we separate the representation of the sentence from that of its context. To this end, for each sentence $s$ of $w$, we note $\Omega_s$ the set of sequences ending with $s$, and whose preceding context is random. The representation of $s$ without context, is, by construction, also the sentence representations of all sequences $w' \in \Omega_s$. Thus, if we denote $y_s^*$ this common representation, the brain response of one subject to a sequence $w'$ can be modeled as

$$\forall w' \in \Omega_s, \quad 🧠(w') = y_s^* + \varepsilon_{w'} ,\quad (2)$$

with $\varepsilon_{w'}$ the context-dependent contribution to 🧠$(w')$. Assuming it is a zero-mean random perturbation we have:

$$\mathbb{E}_{w'}[🧠(w')] = y_s^* ,\quad (3)$$

with $w'$ sampled uniformly in $\Omega_s$. Importantly, we do *not* assume that words are independent of their context but that the *shufflings* defined for each sentence are independent of one another. This statement is true by construction: shuffled contexts are realizations of a uniform sampling of permuted texts. Furthermore, the assumption that activations of shuffled versions of the same context have a zero-mean is not critical: assuming a constant mean would not alter the methods and results, because the final metrics (Pearson correlation) is invariant to such constant.

Similarly, we can retrieve $x_s^*$, the context-independent representation of a particular sequence $s$ in a deep language model

$$\mathbb{E}_{w'}[🤖(w')] = x_s^* .\quad (4)$$

In practice, it is approximated with an average over $K$ *i.i.d.* samples:

$$x_s^* \approx \frac{1}{K} \sum_{k=1}^{K} 🤖(w_k) ,\quad (5)$$

where $w_1, \ldots, w_K$ are sentences uniformly sampled in $\Omega_s$. Given equations (3), (4) and hypothesis *i)*.

$$\bar{y}_s^* = \mathbb{E}_{w'}\left[\overline{🧠(w')}\right]$$
$$= \mathbb{E}_{w'}[f_\theta \circ 🤖(w')]$$
$$\Rightarrow \bar{y}_s^* = f_\theta(x_s^*) ,\quad (6)$$

with $w'$ sampled in $\Omega_s$.

From now on, we note $y^*$ (resp. $x^*$) the context-free representation of the whole story

$w$ extracted from the brain (resp. network) activations. We obtain, $\bar{y}^* = f_\theta(x^*)$.

We now assume that random contexts do not actually affect the brain response to the current sentence in each subject at a given voxel, *i.e.,*

$$ii) \quad y_{|\text{sent}} = y^*_{|\text{sent}} = y^* \ .$$

Under this condition and given equation (6),

$$\overline{y_{|\text{sent}}} = \bar{y}^* = f_\theta(x^*) \ ,$$

and

$$\begin{aligned}\rho\big(y_{|sent}, \overline{y_{|\text{sent}}}\big) &= \rho\big(y^*, f_\theta(x^*)\big) \\ &= \rho\big(y - \varepsilon_w, f_\theta(x^*)\big) \\ &\approx \rho\big(y, f_\theta(x^*)\big) \ ,\end{aligned}$$

with $\varepsilon_w$ the strictly contextual effects in $y$ ($y = y^* + \varepsilon_w$), independent from $x^*$ (context-free).

Finally, under the assumptions that *i*) the deep neural network approximates the average brain response and *ii*) random context is not maintained in memory, the brains-to-brain scores $R = \rho\big(y_{|sent}, \overline{y_{|\text{sent}}}\big)$ are equivalent to the model-to-brain scores $R = \rho\big(y, f_\theta(x^*)\big)$.

## 3 Experiment

To test our model-based approach, we first apply it to the fMRI responses of 75 subjects listening to the same 7 min story analysed in Lerner et al (Nastase et al., 2020)[1]. Thus, for each condition (word, sentence and paragraph), subject and voxel, we compute the model-to-brain correlation $R = \rho\big(y, f_\theta(x^*)\big)$.

The extraction of the fMRI signals $y$, and the estimation of the mapping function $f_\theta$ are standard and thus detailed in Appendices A and C. To estimate context-free representations, we i) scramble the stimulus at the word, sentence or paragraph level, ii) extract the corresponding activations $x$ from a deep language model, and iii) compute $x^*$, as detailed below.

**Scrambling the stimulus at the word, sentence and paragraph level** Words and sentences of the stimulus are delimited using Spacy tokenizer (Honnibal et al., 2020). Note that punctuation marks are not considered as words (*e.g.,* 'time.' forms *one* token, not two). We define paragraphs as contiguous chunks of eight sentences. To 'scramble' a sequence at the word (resp. sentence, paragraph) level, we uniformly shuffle the indices of its words (resp. sentences, paragraphs) and form the new sequence accordingly.

**Extracting deep models' activations** For each version of the scrambled stimulus, we extract the activations from GPT-2 (), a deep neural language model trained to predict a word given its past context. GPT-2 consists of 12 transformer layers of dimensionality 768, 8 heads, and has 1.5 billion parameters in total. We use the model provided by Huggingface (Wolf et al., 2020), trained on a dataset of 8 million web pages.

To extract the activations elicited by a sequence $w$ of $M$ words from layer $l$, we proceed as follows: we tokenize the sequence into subwords called "Byte Pair Encoding" (BPE) (Sennrich et al., 2016) using the GPT-2 tokenizer provided by Huggingface. Then, we feed the network with the $M'$ BPE tokens ($M' \geq M$, up to 256 tokens in memory) and extract the corresponding activations from layer $l$, of shape $(M' \times D)$ with $D = 758$. Then, we sum the activations over the BPEs of each word to obtain a vector of size $(M \times D)$.

All our analyses are based on the eighth layer of GPT-2. We choose GPT-2 because it has been shown to best encode the brain activity elicited by language stimuli (Caucheteux et al., 2021; Schrimpf et al., 2020). We choose its eighth layer because the intermediate layers of transformers have shown to encode relevant linguistic features (Jawahar et al., 2019; Manning et al., 2020) and to better encode brain activity than input and output layers (Caucheteux and King, 2020; Toneva and Wehbe, 2019). Our results successfully generalize to two other architectures as well as to the other intermediate layers of GPT-2 (Appendix F).

**Computing $x^*$ for the word, sentence and paragraph conditions** For each of the word, sentence and paragraph conditions, we compute $x^*$: a context-free representation of $x$. In short, $x^*$ are the activations of GPT-2, averaged over several scrambled contexts. For clarity, we focus on the sentence level to detail the approach.

To build the sentence-level representation

---
[1] http://datasets.datalad.org/?dir=/labs/hasson/narratives

$x^*$ of the stimulus, we use the approximation introduced in equation (5). For each sentence $s$ of one story $w$, we i) generate K=10 sequences ending with $s$, but with scrambled previous context. The scrambled context is uniformly sampled from the other sentences in the same story $w$. Then, ii) we extract the K corresponding activations from GPT-2 (as described in the previous section) and iii) average the activations across the $K$ samples. GPT-2 activations are extracted for each word. Thus, for each of the $M_s$ words of sentence $s$, we obtain a vector $x_s^*$ of shape $M_s \times D$. We concatenate these vectors to obtain $x^*$, a sentence-level representation of the whole story $w$, of shape $M \times D$. This method is adapted from (Caucheteux et al., 2021), in which we computed the average over GPT-2's activations to extract syntactic representations from the input sequence.

**Acoustic features** GPT-2 takes words as input and not sounds. To build $x^*$ at the acoustic level, we simply use non-contextual acoustic features: the word rate ($D = 1$), phoneme rate ($D = 1$) phonemes, stress, and tone (categorical, $D = 117$). For the latter, we use the annotations provided the original Narratives dataset (Nastase et al., 2020).

## 4 Results

The results are displayed in Figure 2B. The hierarchy of temporal receptive fields (TRFs) typically associated with acoustic, word, sentence and paragraph processing along the temporo-parietal axis is remarkably well replicated in both hemispheres (Figure 2B). Notably, both the model-free and model-based methods evidence that the precuneus, the superior frontal gyrus and sulcus are characterized by sentence- and paragraph-level TRFs (Figure 2A and B).

Our results differ from Lerner et al.'s in several ways. First, the acoustic TRFs are slightly more inferior with the model-based method. Second, frontal regions are detected to be associated not only with sentences and paragraphs, but also with words (consistent with (Huth et al., 2016; Caucheteux et al., 2021; Goldstein et al., 2021)). Given that Lerner et al's dataset is not public, it is difficult to quantify these differences and determine whether they reflect an improved sensitivity, or, more simply, inter-individual differences.

Our model-based method can, in principle, be applied to any natural stories. To test this prediction, we extend our analyses to 305 subjects listening to 4.1 hours of fifteen narratives (Figure 2C). Our model-based approach recovers the hierarchy of TRFs, and further reveals additional word- and sentence-level representations in the precuneus and prefrontal regions.

## 5 Discussion

Here, we leverage the modeling power of deep language models to show that the seminal results of Lerner et al. can be retrieved without having subjects listening to multiple scrambled stimuli. Critically, we formalize the assumptions under which 'model-based' and 'model-free' approaches can be linked (Lerner et al., 2011).

Our model-based method recovers the hierarchy of TRFs evidenced by Lerner et al., in the brain of an unusually large cohort of 305 subjects. Thus, our study complements the recent work of (Jain and Huth, 2018; Toneva and Wehbe, 2019; Toneva et al., 2020) who predict brain responses to speech from language models input with variably-long contexts. Specifically, we show that previous model-based results unravel the same mechanisms that was previously identified with model-free approaches.

The replication is not perfect: the acoustic and word TRFs slightly differ between the two methods. This may be explained by individual subject's variability, which is only captured by the model-based approach. Further research, using the non-public data from Lerner et al. should investigate these remaining differences.

In line with previous work (Brennan, 2016; Brennan and Hale, 2019; Gauthier and Levy, 2019; Schrimpf et al., 2020), our study demonstrates that deep neural networks build constructs that predict brain activity, accurately enough to recover the hierarchy of language processing in the brain. The success of replication thus reinforces the idea that naturalistic stimuli and deep neural networks form a powerful couple to study the neural bases of language (Hamilton and Huth, 2020).

# Acknowledgements

This work was supported by the French ANR-20-CHIA-0016 and the European Research Council Starting Grant SLAB ERC-YStG-676943 to AG, and by the French ANR-17-EURE-0017 and the Fyssen Foundation to JRK for his work at PSL.

# Appendix

To replicate Lerner et al.'s findings, we compute the model-to-brain correlation (*cf.* Section 2):

$$R = \rho\big(y, f_\theta(x^*)\big) \ ,$$

for the acoustic, word, sentence and paragraph level respectively. Here, we provide additional details on how to extract the brain signals $y$ and estimate the mapping function $f_\theta$ in order to reproduce the experimental setting used in Section 3.

## A  Brain signals

**Functional MRI dataset**   We use the fMRI recordings of the Narratives dataset (Nastase et al., 2020)[2], a publicly available dataset gathering the brain recordings of 305 subjects listening to narratives. We use the unsmoothed version of the fMRI recordings, already preprocessed in the original dataset. As suggested in the original paper, we reject subject / narrative pairs because of noisy recordings, resulting in 617 unique (story, subject) pairs and 4.1 hours of audio stimulus in total. To replicate the results of Lerner et al. (2011), we restrict the analyses to the 75 subjects listening to the 'Pieman' story (7 min long), including the seven subjects analysed in the original paper (only the data for non-scrambled stimuli are publicly available). Then, we extend the analyses to the brain recordings of 305 subjects listening to fifteen narratives (from 3 min to 57 min), from the same dataset (Nastase et al., 2020). For both analyses, we only have access and thus use the brain recordings elicited by regular –i.e non scrambled– version of the stimuli.

## B  Encoding features

**Deep language models' activations**   In Section 3, we extract the activations of GPT-2 (🤖), a deep neural language model trained to predict a word given its past context. It consists of 12 transformer layers of dimensionality 768, 8 heads, and has 1.5 billion parameters in total. We use the model provided by Huggingface (Wolf et al., 2020), trained on a dataset of 8 million web pages.

---

[2]http://datasets.datalad.org/?dir=/labs/hasson/narratives

To extract the activations elicited by a sequence $w$ of $M$ words from a layer $l$, we proceed as follows: we tokenize the sequence into subwords called "Byte Pair Encoding" (BPE) (Sennrich et al., 2016) using the GPT-2 tokenizer provided by Huggingface. Then, we feed the network with the $M'$ BPE tokens ($M' \geq M$, up to 256 tokens in memory) and extract the corresponding activations from layer $l$, of shape $(M' \times D)$ with $D = 758$. Then, we sum the activations over the BPEs of each word to obtain a vector of size $(M \times D)$.

All our analyses are based on the eighth layer of GPT-2. We choose GPT-2 because it has been shown to best encode the brain activity elicited by language stimuli (Schrimpf et al., 2020). We choose its eighth layer because the intermediate layers of transformers have shown to encode relevant linguistic features (Jawahar et al., 2019; Manning et al., 2020) and to better encode brain activity than input and output layers (Caucheteux and King, 2020; Toneva and Wehbe, 2019).

**Scrambling the stimulus at the word, sentence and paragraph level**   Words and sentences of the stimulus are delimited using Spacy tokenizer (Honnibal et al., 2020). Note that punctuation marks are not considered as words (*e.g.,* 'time.' forms *one* token, not two). We define paragraphs as contiguous chunks of eight sentences. To 'scramble' a sequence at the word (resp. sentence, paragraph) level, we uniformly shuffle the indices of its words (resp. sentences, paragraphs) and form the new sequence accordingly.

**Computation of $x^*$ for the word, sentence and paragraph conditions**   In Section 2, we compute a context-free representation $x^*$ for the word, sentence and paragraph condition. In short, $x^*$ are the activations of GPT-2, averaged over several scrambled contexts. For clarity, we focus on the sentence level to detail the approach. To build the sentence-level representation $x^*$ of the stimulus, we use the approximation introduced in equation (5). For each sentence $s$ of one story $w$, we i) generate K=10 sequences ending with $s$, but with scrambled previous context. The scrambled context is uniformly sampled from the other sentences in the same story $w$. Then, ii) we extract the K corresponding activations from GPT-2 (as described in the previous section) and iii) average the activations across the $K$ samples. GPT-2 activations are extracted for each word. Thus, for each of the $M_s$ words of sentence $s$, we obtain a vector $x_s^*$ of shape $M_s \times D$. We concatenate these vectors to obtain $x^*$, a sentence-level representation of the whole story $w$, of shape $M \times D$. This method is adapted from (Caucheteux et al., 2021), in which the authors compute the average over GPT-2 activations to extract syntactic representations from the input sequence.

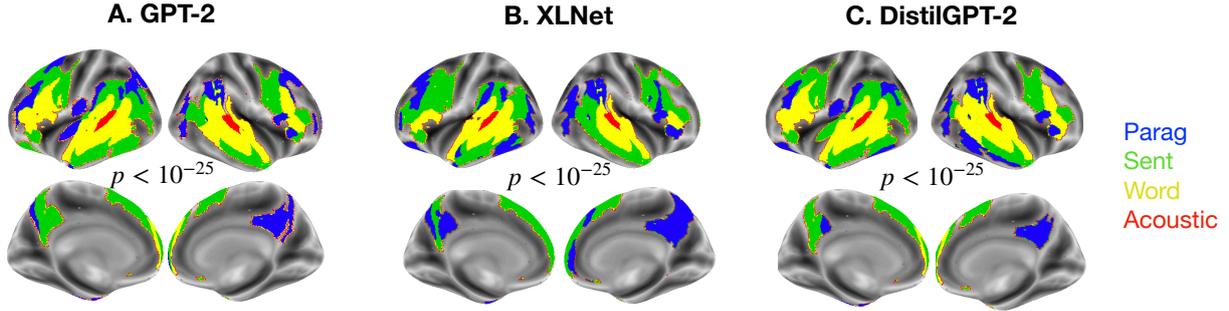

Figure 3: **Replication to two other architectures.** Same as Figure 2.C but using the intermediate layers of XLNet and Distilgpt2 causal architectures ($l = 4$ for Distilgpt2, out of 6 layers in total and $l = 8$ for XLNet, out of 12 layers in total). As in Figure 2.C, the significance threshold is set to $p < 10^{-25}$.

**Acoustic features** GPT-2 takes words as input and not sounds. To build $x^*$ at the acoustic level, we simply use non-contextual acoustic features: the word rate ($D = 1$), phoneme rate ($D = 1$) phonemes, stress, and tone (categorical, $D = 117$). For the latter, we use the annotations provided the original Narratives dataset (Nastase et al., 2020).

## C  Mapping $x^*$ onto the brain

The linear function $f_\theta$ maps $x^*$ onto $y$, the fMRI recordings of one subject at one voxel. Vector $y$ is of length $T$, the number of fMRI time samples, whereas $x^*$ is of length $M$, the number of words (or phonemes for acoustic features) in the story. To align the two time domains, we apply the function $g : \mathbb{R}^{M \times D} \mapsto \mathbb{R}^{T \times 5D}$ that i) sums the features $x^*$ between the successive fMRI time samples, and ii) uses a Finite-Impulse Response model (FIR) with five delays. Thus, $f_\theta = f'_\theta \circ g$, with $f'_\theta$ a linear function whose parameters $\theta$ are learned, and $g$ a temporal alignment function.

To estimate $\theta$, we fit an $\ell_2$-penalized linear regression to predict $y$ given $g(x^*)$ on a training set of time samples. $\theta$ thus minimizes

$$\underset{\theta' \in \Theta}{\mathrm{argmin}} \ \|y_{\mathrm{train}} - f_{\theta'} \circ g(x^*_{\mathrm{train}})\|^2 + \lambda \|\theta'\|^2 \ ,$$

with $\lambda$ the regularization parameter. We assess the mapping with a Pearson correlation score evaluated on the left out times samples:

$$R = \rho\Big(y_{\mathrm{test}}, f_\theta \circ g(x^*_{\mathrm{test}})\Big) \ .$$

In practice, $x^*$ and $g(x^*)$ are standardized (0-mean, 1-std) and brain signals $y$ are scaled based on quantiles using scikit-learn RobustScaler (Pedregosa et al., 2011) with quantile range (.01, .99). We use the RidgeCV implementation of scikit-learn with a pool of twenty possible penalization parameters between $10^{-3}$ and $10^6$. We learn $f_\theta$ on 90% of the $T$ time samples, and compute the correlation scores $R$ on the 10% left out data. We repeat the procedure on 10 test folds using a cross-validation setting, following the KFold implementation of scikit-learn without shuffling. Finally, we average the $R$ over the 10 folds to obtain one model-to-brain correlation score per subject, voxel and feature space $x^*$.

## D  Brain parcellation

In Figure 2, we use a subdivision of Destrieux' atlas (Destrieux et al., 2010). Regions of more than 200 vertices are split into smaller regions, so that each region contains at most 200 vertices. Thus, from the 75 regions of Destrieux' atlas (in each hemisphere), we obtain a parcellation of 465 brain regions per hemisphere.

## E Significance

In Figure 2, we test whether the model-to-brain correlations ($R$) are significantly different from zero. To this aim, we use a two-sided Wilcoxon test across subjects ($N = 75$ in Figure 2B, $N = 305$ in Figure 2A), corrected using False Discovery Rate (FDR) across the 465 region of interests in each hemisphere.

## F Generalization to other transformer architectures

In Figure 3 (B and C), we replicate our results (Figure 2.C) on the activations of two other causal transformer architectures: XLNet (Yang et al., 2020) and Distilgpt2 (Figure 3.C), using the implementation from Huggingface[3].

---

[3] https://huggingface.co/